\documentclass[aps,pre,showpacs,showkeys,square,numbers,amssymb,amsmath,nobibnotes,nofootinbib,superscriptaddress]{revtex4}
\usepackage{docs}%
\usepackage{amsmath}
\usepackage{amsfonts}
\usepackage{amssymb}
\usepackage{verbatim}
\usepackage{epsfig}
\usepackage{eucal}
\usepackage{graphicx}
\usepackage[sort&compress]{natbib}

\newtheorem{conj}{Conjecture}
\newtheorem{Example}{Example}

\def\Tr{\mathrm{Tr}}
\def\Det{{\rm Det}}
\def\S{{\mathfrak  S}}
\def\Q{{\mathbb Q}}
\def\tK{{\widetilde K}}
\begin{document}


\title{Nonlinear Random Matrix Statistics, symmetric functions and hyperdeterminants.}


\author{Jean-Gabriel Luque}
\affiliation{LITIS - D\'epartement d'informatique de
l'Universit\'e de Rouen.\\ Avenue de l'Universit\'e - BP 8 76801
Saint Etienne DU Rouvray, France}

\author{Pierpaolo Vivo}
\affiliation{ICTP - Abdus Salam International Centre for Theoretical Physics\\
Strada Costiera 11, 34014 Trieste, Italy}

\date{\today}

\begin{abstract}
Nonlinear statistics (i.e. statistics of permanents) on the eigenvalues of invariant random matrix models are considered for the three Dyson's symmetry classes $\beta=1,2,4$.
General formulas in terms of hyperdeterminants are found for $\beta=2$. For specific cases and all $\beta$s, more computationally efficient results are obtained, based on symmetric functions expansions.
As an application, we consider the case of quantum transport in chaotic cavities
extending results from [D.V. Savin, H.-J. Sommers and W. Wieczorek, {\it Phys. Rev. B} {\bf 77}, 125332 (2008)].
\end{abstract}
\pacs{73.23.-b, 02.10.Yn, 24.60.-k, 73.63.Kv}
\keywords{Random Matrix, Nonlinear statistics, Chaotic cavity, Selberg integral, hyperdeterminants, symmetric functions.}
\maketitle

\section{Introduction}

Random matrices are known to find applications in many physical systems \cite{mehtabook,guhr}. In the present work, we focus on rotationally invariant ensembles of
$N\times N$ matrices $\mathbf{H}$ (symmetric, hermitian or quaternion self-dual), for which the joint probability density
(jpd) of the $N$ real eigenvalues $\{T_i\}$ can be generically written as:
\begin{equation}\label{jpd}
P(T_1,\ldots,T_N)=\frac{1}{Z_\omega(\beta,N)}\prod_{j<k}|T_j-T_k|^\beta\prod_{i=1}^N \omega(T_i)
\end{equation}
where $\beta$ is the Dyson index of the ensemble ($\beta=1,2,4$ respectively), $\omega(x)$ a certain weight function
and $Z_\omega(\beta,N)$ is the normalization constant.
The classical ensembles of random matrix theory correspond to the following weight functions:
\begin{align}
\omega(x) &=e^{-x^2/2} &\qquad -\infty<x<\infty &\qquad \mbox{Gaussian Ensemble} &\quad (\textsf{GXE})\\
\omega(x) &=e^{-x}x^{\alpha -1} &\qquad x>0 &\qquad \mbox{Laguerre Ensemble}  &\quad (\textsf{LXE})\\
\label{jacweight}\omega(x) &=x^{\alpha-1} (1-x)^{\gamma-1} &\qquad 0<x<1 &\qquad \mbox{Jacobi Ensemble} &\quad  (\textsf{JXE})
\end{align}

where $\mathsf{X}=\{\mathsf{O},\mathsf{U},\mathsf{S}\}$ stands for Orthogonal, Unitary and Symplectic ($\beta=1,2,4$ respectively). The normalization constant
$Z_\omega(\beta,N)$ can be computed via the celebrated Selberg's integral:
\begin{align}\label{selb}
\nonumber S_n(a,b,c) &:=\int_0^1\cdots\int_0^1 dt_1\cdots dt_n\prod_{i=1}^n t_i^{a-1}(1-t_i)^{b-1}\prod_{1\leq i<j\leq n}|t_i-t_j|^{2c}\\
&=\prod_{j=0}^{n-1}\frac{\Gamma(a+jc)\Gamma(b+jc)\Gamma(1+(j+1)c)}{\Gamma(a+b+(n+j-1)c)\Gamma(c+1)}
\end{align}
and its generalizations. In particular, we have:
\begin{align}
Z_{\omega\equiv G}(\beta,N) &= (2\pi)^{N/2}\prod_{j=1}^N \frac{\Gamma\left(1+\frac{\beta}{2}j\right)}{\Gamma\left(1+\frac{\beta}{2}\right)}\\
Z_{\omega\equiv L}(\beta,N) &= \prod_{j=0}^{N-1}\frac{\Gamma(\alpha+ j\beta/2)\Gamma((j+1)\beta/2)}{\Gamma(\beta/2)}\\
Z_{\omega\equiv J}(\beta,N) &= S_N(\alpha,\gamma,\beta/2)
\end{align}

In many physical applications, one is interested in so-called \emph{linear statistics} on the $N$ eigenvalues, i.e. random variables of the form:
\begin{equation}\label{linearstat}
\mathcal{A} =\sum_{i=1}^N f(T_i)
\end{equation}
where the function $f(x)$ may well be highly non-linear (see \cite{chen,politzer,vivoPRL,forresterjp} and references therein for physical applications).
In particular, general methods are available \cite{vivoPRL,beenakkerPRB} to compute in principle the mean and variance of any linear statistics from a generic invariant ensemble,
at least in the large $N$ limit.

Conversely, much less is known for \emph{nonlinear} statistics, i.e. functions involving products of different eigenvalues (see however \cite{lytova,giraud}). One may for example consider the following
random variable:
\begin{equation}\label{permanent}
\mathcal{T}_{\mathbf{\Psi}} =\mathrm{perm}(\mathbf{\Psi}):=\sum_{\pi\in\S_N}\prod_{i=1}^N \psi_{\pi(i)}(T_i)
\end{equation}
where $\mathrm{perm}$ stands for the \emph{permanent} of the $N\times N$ matrix $\mathbf{\Psi}=(\psi_i(T_j))_{1\leq i,j\leq N}$,
the sum runs over the permutations $\pi$ of the first $N$ integers ($\S_N$ is the symmetric group), and $\{\psi_i(x)\}$ is a set of $N$ given functions. Clearly, the general definition above:
\begin{enumerate}
\item is invariant under permutations of the $T_i$;
\item incorporates as special cases e.g. powers of the determinant of
$\mathbf{H}$, ($(\det\mathbf{H})^\kappa$) \cite{mehtajp,cicuta} [eq. \eqref{permanent} when $\psi_i(x)=x^\kappa\quad\forall i$]
as well as traces of higher powers of $\mathbf{H}$, ($\mathrm{Tr}\mathbf{H}^\kappa$) [eq. \eqref{permanent} when $\psi_i(x)=x^\kappa\mbox{  for  }i=1\mbox{  and  }1\mbox{  otherwise}$].
\end{enumerate}

The aim of this paper is to study the statistics of permanents $\mathcal{T}_{\mathbf{\Psi}} $ on classical random matrix ensembles.
Our motivation comes from the problem of quantum transport in open chaotic cavities supporting
$N_1$ and $N_2$ electronic channels in the two attached leads. A detailed account
of the problem and its link with the Jacobi ensemble of random matrices is provided in Appendix \ref{AppA}.
Our more general approach allows to extend results from \cite{savin} in a clear and computationally efficient way.

We base our analysis on the theories of \emph{hyperdeterminants}
and \emph{symmetric functions}. The former is detailed in Appendix
\ref{AppB} and involves multidimensional generalizations of  the
conventional determinant: it provides a very general (although not
always efficient) way to write averages of permanents as sums of
determinants for $\beta=2$. The latter is detailed in Appendix
\ref{AppC} and will be used to produce less general, but quite
powerful formulae for a few physically interesting cases. The use
of hyperdeterminants and symmetric functions in random matrix
contexts is not new (see e.g.
\cite{savinnew,novaes,forrselberg,kanzake,fyodk} and references
therein). Here we apply similar methods to a different problem.
Note that the links between Selberg integrals and
hyperdeterminants have been already investigated by one of the
author with Jean-Yves Thibon \cite{luquethibon,luqueiop}.

The plan of this paper is as follows. In section \ref{hyper} we provide a general hyperdeterminant formula for the average of permanents valid for $\beta=2$, any $\mathbf{\Psi}$
and any benign weight $\omega(x)$.
While being very general, the practical implementation becomes rapidly unwieldly due to an exponential growth of the number of terms with $N$. In section \ref{symm},
we resolve this efficiency issue adopting symmetric functions expansions. The resulting formulae do \emph{not} increase in complexity when $N$ grows, thus making the
numerical implementation extremely efficient, though at the price of a loss in generality. These improved formulae are valid for $\beta=1,2,4$ for the Jacobi weight with $\gamma=1$ and we restrict ourselves to the most interesting case $\psi_i(x)=x^{\lambda_i}$. Finally in section \ref{conclusions} we offer some concluding remarks. In the appendices, we give a detailed account
of the problem of quantum transport in chaotic cavities which constitutes our motivation (\ref{AppA}) and some remarks abour hyperdeterminants (\ref{AppB}) and symmetric functions (\ref{AppC}). In
(\ref{AppD}), guided by the numerics, we perform an asymptotic analysis for large $N$ and put forward a \emph{factorization} conjecture that will be studied in more details in a forthcoming paper \cite{vivopreparation}.

\section{Hyperdeterminant formula for statistics of permanents for $\beta=2$}\label{hyper}

We are now interested in computing the following average:
\begin{equation}\label{permaverage}
\langle \mathrm{perm}(\mathbf{\Psi})\rangle =\frac{1}{Z_\omega(\beta,N)}\int\prod_{j=1}^N d T_j\ \omega(T_j)\prod_{j<k}|T_j-T_k|^\beta \mathrm{perm}(\psi_i(T_j))_{1\leq i,j\leq N}
\end{equation}
where $\omega(x)$ is one of the classical random matrix weights and the integrals run over the appropriate support. Hereafter it is assumed that both the measure $\omega(x)$ and the functions
$\psi_i(x)$ are benign, i.e. they ensure existence and convergence of the integrals involved.
\begin{Example}
Consider $\psi_i(x)=x^{\lambda_i}$, where $\lambda=[\lambda_1,\lambda_2,\ldots ,\lambda_N]$ is a decreasing partition $(\lambda_1\geq\lambda_2\geq\ldots\geq\lambda_N)$. Then:
\begin{equation}
\langle \mathrm{perm}(\mathbf{\Psi})\rangle = N!\ \langle T_1^{\lambda_1}\cdots T_N^{\lambda_N}\rangle
\end{equation}
\end{Example}

Definition \eqref{permaverage} is very general: it requires to specify the Dyson index $\beta$, the measure $\omega(x)$ and the set of functions $\{\psi_i(x)\}$.
In this section, we will focus mainly on the unitary case ($\beta=2$) , all the remaining 'degrees of freedom' being left untouched. In principle, the same reasoning could be applied to
any even $\beta$, but the resulting formulae are too complicated for any practical use.

The main technical tools are the following:
\begin{itemize}
\item The expansion of a hyperdeterminant as a sum of conventional
determinants (see eq. \eqref{GegenExp} in Appendix \ref{AppB}).
\item A generalization of Heine's theorem for determinants
\footnote{The original Heine's theorem can be found in
\cite{heine}. Conversely, eq. \eqref{heine} does not appear
explicitly in literature but a version for totaly alternated
hyperdeterminants can be found in  \cite{boussicault} and its
proof is straightforward.}. Given $\ell$ sets of $N$ functions $f_i^{(s)}(x)$
($s=1,\ldots,\ell$ and $i=1,\ldots, N$) and a benign integration
measure $\mu(x)$ the following holds:
\begin{align}\label{heine}
\nonumber\int\cdots\int\prod_{i=1}^N d\mu(x_i) &\prod_{s=1}^k\mathrm{perm}(f_i^{(s)}(x_j))_{1\leq i,j\leq N}\prod_{s=k+1}^\ell\det(f_i^{(s)}(x_j))_{1\leq i,j\leq N}
=\\
&=N!\ \Det_{\{k+1,\ldots,\ell\}}\left(\int d\mu(x) \prod_{s=1}^\ell f_{i_s}^{(s)}(x)\right)_{1\leq i_1,\ldots, i_\ell\leq N}
\end{align}
\item The Cauchy's double alternant evaluation \cite{kratt}. For any pair of vectors $\mathbf{X}=\{X_1,\ldots,X_n\}$ and $\mathbf{Y}=\{Y_1,\ldots,Y_n\}$
the following holds
\begin{equation}\label{identityS}
\det\left(\frac{1}{X_i+Y_j}\right)_{1\leq i,j\leq n}=\frac{\prod_{1\leq i<j\leq n}(X_i-X_j)(Y_i-Y_j)}{\prod_{1\leq i,j\leq n}(X_i+Y_j)}
\end{equation}
\end{itemize}


Let us introduce for $\beta=2$ a special case of hyperdeterminant $\Det_+$ (see the general definition and properties in Appendix \ref{AppB}) of the multi-indices tensor $M$
defined by
\begin{equation}
\Det_+ (M_{i,j,k})_{1\leq i,j,k\leq
N}=\frac1{N!}\sum_{\sigma_1,\sigma_2,\sigma_3\in\S_N^3}\epsilon(\sigma_2\sigma_3)\prod_{i=1}^N M_{\sigma_1(i),\sigma_2(i),\sigma_3(i)}.
\end{equation}
where the sum runs over permutations $\sigma_1,\sigma_2,\sigma_3$ of the first $N$ integers and $\epsilon$ is the product of their signatures.

In the following, we will make use of two of the properties just stated above, namely:
\begin{enumerate}
\item The expansion in terms of conventional determinants (see again Appendix \ref{AppB})
\begin{equation}\label{eq2}
 \Det_+ (M_{i,j,k})_{1\leq i,j,k\leq
N}=\sum_{\sigma\in\S_N} \det(M_{\sigma(i),i,j}).
\end{equation}
\item
The generalization of the Heine theorem for hyperdeterminants \eqref{heine}:
\begin{align}\label{eq1}
\nonumber\int\cdots\int {\rm perm}(f_i(x_j))_{1\leq i,j\leq N}\det(g_i(x_j))_{1\leq i,j\leq N} &\det(h_i(x_j))_{1\leq i,j\leq N}\prod_{j=1}^N \omega(x_j)dx_j=\\
&=N!\ {\rm
Det}_+\left(\int dx\ \omega(x) f_i(x)g_j(x)h_k(x)\right)_{1\leq i,j,k\leq N}.
\end{align}
\end{enumerate}
From Heine's theorem, the evaluation of the average \eqref{permaverage} for $\beta=2$ is quite straightforward.
Noting that:
\begin{equation}
\prod_{j<k}(T_j-T_k)^2 = \det(T_i^{j-1})^2_{1\leq i,j\leq N}
\end{equation}
one simply has:
\begin{equation}
\langle \mathrm{perm}(\mathbf{\Psi})\rangle =\frac{N!}{\ Z_\omega(2,N)} \Det_+ \left(\int dx\ \omega(x) \psi_i(x) x^{j+k-2}\right)_{1\leq i,j,k\leq N}
\end{equation}

From the first property stated above, this can be expanded as a sum over the
symmetric group $\S_N$

\begin{equation}\label{finalbeta2hyper}
\boxed{\langle \mathrm{perm}(\mathbf{\Psi})\rangle =\frac{N!}{ Z_\omega(2,N)} \sum_{\sigma\in\S_N}\det\left(\int dx\ \omega(x) \psi_{\sigma(i)}(x) x^{i+j-2}\right)_{1\leq i,j\leq N}}
\end{equation}

Eq. \eqref{finalbeta2hyper} is the main result of this section. It
expresses in full generality the average of any permanent for any
invariant ensemble with $\beta=2$ as a sum of $N!$ determinants.
Clearly, due to the exponential growth with $N$ of its complexity,
formula \eqref{finalbeta2hyper} is only practical when $N<10$. We
will offer in section \ref{symm} a less general, but more powerful
way to compute the sought average. 

\begin{Example}
Suppose we take the Jacobi weight with $\gamma=1$, $\omega(x)=x^{\alpha-1}$ (relevant for the quantum transport problem) and the nonlinear statistics $\psi_i(x)=x^{\lambda_i}$, where
$\lambda=[\lambda_1,\dots,\lambda_N]$
with $\lambda_1\geq\dots\geq\lambda_N$. Then formula \eqref{finalbeta2hyper} reads:
\begin{equation}\label{momentshyper}
\langle T_1^{\lambda_1}\cdots T_N^{\lambda_N}\rangle_\alpha = \frac{\langle \mathrm{perm}(\mathbf{\Psi})\rangle}{N!}=
\frac{ \sum_{\sigma\in\S_N}\det\left(\int_0^1 dx\ x^{\alpha+\lambda_{\sigma(i)}+i+j-3}\right)_{1\leq i,j\leq N}}{ Z_{\omega\equiv J}(2,N)}=
\frac{\sum_{\sigma\in\S_N}\det\left(\frac{1}{\alpha+\lambda_{\sigma(i)}+i+j-2}\right)_{1\leq i,j\leq N}}{ Z_{\omega\equiv J}(2,N)}
\end{equation}
Further simplifications are achieved employing the Cauchy's double alternant identity, which finally yields:
\begin{equation}\label{avpermJac}
\langle T_1^{\lambda_1}\cdots T_N^{\lambda_N}\rangle_\alpha=\frac{\prod_{i<j}(i-j)}{ Z_{\omega\equiv J}(2,N)}
\sum_{\sigma\in\S_N} \frac{\prod_{1\leq i<j\leq N}(\lambda_{\sigma(i)}+i-\lambda_{\sigma(j)}-j)}{\prod_{i,j=1}^N (\lambda_{\sigma(i)}+i+j+\alpha-2)}
\end{equation}
Eq. \eqref{avpermJac} extends in a compact form the results from \cite{savin} to arbitrary values of $\lambda_j$.
In the special case $\lambda=(1,\ldots,1)$, we note that \eqref{avpermJac} perfectly matches the value for the average of the determinant for a Jacobi ensemble with $\gamma=1$, as
computed from the Selberg integral \eqref{selb}:
\begin{equation}
\langle\det(\mathbf{H})\rangle_\alpha=\langle T_1\cdots T_N\rangle_\alpha=\frac{S_N(\alpha+1,1,1)}{S_N(\alpha,1,1)}=\prod_{j=0}^{N-1}\frac{\alpha+j}{\alpha+N+j}
\end{equation}
thanks to the following (easy to prove) identity valid $\forall\alpha,n$:
\begin{equation}\label{identity}
\frac{n!\ \prod_{i<j}(i-j)^2}{\prod_{i,j=1}^n (i+j+\alpha-1)}=
\prod_{j=0}^{n-1}\frac{\Gamma(\alpha+j+1)\Gamma(j+1)\Gamma(j+2)}{\Gamma(\alpha+1+n+j)}
\end{equation}
\end{Example}

\section{More efficient symmetric function expansions for the Jacobi weight}\label{symm}
In this section, we are able to provide more efficient and user-friendly formulae for a special nonlinear statistics, namely the quantity:
\begin{equation}\label{quantity}
\langle T_1^{\lambda_1}\cdots T_N^{\lambda_N}\rangle_\alpha =\frac{1}{Z_{\omega\equiv J}(\beta,N)}\int_{[0,1]^N}d T_1\cdots dT_N \
T_1^{\lambda_1}\cdots T_N^{\lambda_N}\prod_{j<k}|T_j-T_k|^\beta\prod_{i=1}^N T_i ^{\alpha-1}
\end{equation}
where the average is taken with respect to the Jacobi weight \eqref{jacweight} with $\gamma=1$ and $\beta=1,2,4$.

Such object is of interest for the statistics of moments of experimental observables in the problem of quantum transport in ballistic chaotic cavities. A detailed overview of the problem
is provided in Appendix \ref{AppA}. It is likely that the method we present here, based on symmetric function expansions, may be applied with slight modifications to a number of other measures
and observables.

The main tools are the following:
\begin{itemize}
\item The following identity (hereafter we use the notation of \cite{macdonald}):
\begin{equation}\label{identityperm}
\mathrm{perm}(T_i^{\lambda_j})_{1\leq i,j\leq N}=\lambda^{!}m_\lambda
\end{equation}
where $m_\lambda$ is the \emph{monomial symmetric function} \cite{macdonald}
\begin{equation}\label{monomialdef}
m_\lambda := m_\lambda(T_1,\ldots, T_N) =\sum_{I}
T_1^{I_{1}}\cdots T_N^{I_N}
\end{equation}
summed over all distinct permutations $I$ of $\lambda$,
and $\lambda^!=n_0!\dots n_k!\dots$ if $n_i$ denotes the number of
occurences of $i$ in $\lambda$ (for example,
$[5,5,5,3,3,2,1,0,0,0]^!=3!0!2!1!1!3!$). \item The well-known link
between Selberg-type integrals and Jack polynomials given by the
Kadell formula \cite{kadell}:
\begin{align}
\nonumber I_\lambda^{(\frac{1}{c})} &:=\int_{[0,1]^N}
J_\lambda^{\left(\frac1c\right)}(T_1,\dots, T_N)
\prod_{i<
j}|T_i-T_j|^{2c}\prod_{i=1}^N(1-T_i)^{b-c(N-1)-1}T_i^{a-c(N-1)-1}dT_i\\
&=J_\lambda^{\left(\frac1c\right)}(1,\dots,1)\prod_{i=1}^N\frac{\Gamma(ci+1)\Gamma(b+c(i+1))\Gamma(\lambda_i+a+c(1-i))
}{ \Gamma(c+1)\Gamma(\lambda_i+a+b+c(1-i))}.\label{kadell}
\end{align}

where the value of $J_{\lambda}^{\left(\frac1c\right)}(1,\dots,1)$
is known to be
\begin{equation}
J_{\lambda}^{\left(\xi\right)}(1,\dots,1)=\xi^{|\lambda|}\prod_{i=1}^N{\Gamma(\frac1\xi(N-i+1)\lambda_i)\over\Gamma(\frac1\xi(N-i+1))}
\end{equation}
if $N\geq \ell(\lambda)$ (where $\ell(\lambda)$ denotes the \emph{length} of the partition $\lambda$ and $|\lambda|$ the sum of its nonzero parts) and $0$ otherwise \cite{macdonald}. The
Jack polynomials for $c=1$ are proportional to Schur functions and for
$c=1/2$ are known as zonal polynomials (see Appendix \ref{AppC} for details).
\end{itemize}

We first illustrate in some detail the method for $\beta=2$ ($c=1$) based on Schur function expansions, where we provide also a detailed asymptotic analysis for $N\to\infty$, and then the cases $\beta=1,4$ together on the same footing.

\subsection{The Unitary case ($\beta=2$)}

Consider the expansion of $m_\lambda$ in the Schur basis
\cite{macdonald}
\begin{equation}
 m_\lambda=\sum_{\mu} \tK_\lambda^\mu s_\mu.
\end{equation}
The coefficients $\tK_\lambda^\mu$ are obtained inverting the Kostka matrix \cite{macdonald}, an operation that can easily be performed
by symbolic computation routines.

Replacing the permanent by its expansion in the Schur basis one
finds (for $\alpha=1$):
\begin{equation}\label{instr1}
 \langle T_1^{\lambda_1}\dots
 T_N^{\lambda_N}\rangle_{\alpha=1}=\frac{\lambda^!}{N!\ Z_{\omega\equiv J}(2,N)}\sum_\mu   \tK_\lambda^\mu \int_{[0,1]^N}
s_\mu(T_1,\dots,T_N)\det(T_i^{j-1})_{1\leq i,j\leq N}^2\prod_{j=1}^N dT_j
\end{equation}
The integral on the right-hand side is readily recognized as a special case of Kadell's integral \eqref{kadell} for $c=1$, leading immediately (after simplifications) to the final result
(eq. \eqref{mainjacobibeta11} and \eqref{mainjacobibeta2}). However, it is more instructive to proceed directly from \eqref{instr1} and notice that
each Schur function is itself the quotient of two determinants \cite{macdonald}:
\begin{equation}
 s_\mu(T_1,\dots, T_N)={\det(T_i^{\mu_{j}+N-j})_{1\leq i,j\leq N}\over
 \det(T_i^{j-1})_{1\leq i,j\leq N}}.
\end{equation}
It follows that the integral in \eqref{instr1} becomes:
\begin{equation}
 \langle T_1^{\lambda_1}\dots
 T_N^{\lambda_N}\rangle_{\alpha=1}=\frac{\lambda^!}{N!\ Z_{\omega\equiv J}(2,N)} \sum_\mu \tK_\lambda^\mu\int_{[0,1]^N}
\det(T_i^{\mu_{j}+N-j})_{1\leq i,j\leq N} \det(T_i^{j-1})_{1\leq
i,j\leq N}\prod_{j=1}^N dT_j
\end{equation}
and using the Heine
theorem, each multiple integral in the sum can be again converted into a determinant. This procedure can be easily implemented for weight functions different from Jacobi, and
will lead to general expressions for averages like $ \langle T_1^{\lambda_1}\dots
 T_N^{\lambda_N}\rangle$ as computationally efficient sums of determinants. In the present case, following one strategy or another and exploiting the Cauchy's identity \eqref{identityS} we obtain:

\begin{equation}\label{mainjacobibeta11}
\langle T_1^{\lambda_1}\dots
 T_N^{\lambda_N}\rangle=\lambda^!{\prod_{i<j}(i-j)\over {Z_{\omega\equiv
J }(2,N)}}\sum_\mu \tK_\lambda^\mu
 {\prod_{i<j}(\mu_{i}-\mu_{j}+j-i)\over\prod_{i,j}(\mu_{i}+N-i+j)}.
\end{equation}

The parameter $\alpha$ can be introduced easily:
\begin{equation}\label{mainjacobibeta2}
\boxed{\langle T_1^{\lambda_1}\dots
 T_N^{\lambda_N}\rangle_\alpha=\lambda^!{\prod_{i<j}(i-j)\over {Z_{\omega\equiv J }(2,N)}}\sum_\mu \tK_\lambda^\mu
 {\prod_{i<j}(\mu_{i}-\mu_{j}+j-i)\over\prod_{i,j}(\mu_{i}+N-i+j+\alpha-1)}}.
\end{equation}

The formula \eqref{mainjacobibeta2} is the main result of this section. It provides a very efficient algorithm (see examples below), since the size
of the sum does not depend on the size $N$ of the alphabet (compare it with formula \eqref{avpermJac}). Knowing
the Schur expansion of the monomial, the computation is immediate.
Let us illustrate this method on the following example.
\begin{Example}
Let $\lambda=[4,3,2]$. One has
\[\begin{array}{rcl}
m_{[4,3,2]}&=&s_{[4,3,2]}-s_{[4,3,1,1]}-s_{[4,2,2,1]}+2\,s_{[4,2,1,1,1]}\\
&&-2\,s_{[4,1,
1,1,1,1]}-2\,s_{[3,3,3]}+s_{[3,3,2,1]}-2\,s_{[3,2,1,1,1,1]}\\
&&+4\,s_{[3,1
,1,1,1,1,1]}+2\,s_{[2,2,1,1,1,1,1]}\\
&&-6\,s_{[2,1,1,1,1,1,1,1]}+6\,s_{[1, 1,1,1,1,1,1,1,1]}
\end{array}\]

Hence,
\[
\begin{array}{rcl}
\langle T_1^4T_2^3T_3^2\rangle_\alpha &=&f_{[4,3,2]}-f_{[4,3,1,1]}-f_{[4,2,2,1]}+2\,f_{[4,2,1,1,1]}\\
&&-2\,f_{[4,1,
1,1,1,1]}-2\,f_{[3,3,3]}+f_{[3,3,2,1]}-2\,f_{[3,2,1,1,1,1]}\\
&&+4\,f_{[3,1
,1,1,1,1,1]}+2\,f_{[2,2,1,1,1,1,1]}\\
&&-6\,f_{[2,1,1,1,1,1,1,1]}+6\,f_{[1, 1,1,1,1,1,1,1,1]}
\end{array}
\]
where
\begin{equation*}
f_\mu=\lambda^!{\prod_{i<j}(i-j)\over {Z_{\omega\equiv
J}(2,N)}}{\prod_{i<j}(\mu_{i}-\mu_{j}+j-i)\over\prod_{i,j}(\mu_{i}+N-i+j+\alpha-1)}
\end{equation*}
 if the length of $\mu$ is less or equal to the number $N$ of
 variables $T_i$ and $0$ otherwise.
 \begin{itemize}
\item For $N=3$, one finds after simplifications
\[
 \langle T_1^4T_2^3T_3^2\rangle_\alpha={\frac {\left (28+10\,\alpha+{\alpha}^{2}\right )\left (2+\alpha
\right )^{2}\left (3+\alpha\right )\alpha\,\left (1+\alpha\right
)^{2} }{\left (6+\alpha\right )^{2}\left (4+\alpha\right )\left
(7+\alpha \right )\left (5+\alpha\right )^{3}\left (8+\alpha\right
)}}
\]
\item For $N=8$

 {\footnotesize\[  \langle T_1^4T_2^3T_3^2\rangle_\alpha={\frac {P(\alpha )\left (5+\alpha\right
)\left (6+\alpha\right ) \left (7+\alpha\right )}{\left
(11+\alpha\right )\left (9+\alpha \right )\left (12+\alpha\right
)\left (10+\alpha\right )\left (13+ \alpha\right )\left
(14+\alpha\right )^{2}\left (15+\alpha\right )^{3} \left
(16+\alpha\right )^{2}\left (17+\alpha\right )\left (18+\alpha
\right )}}.
\]
} with {\footnotesize
\[\begin{array}{l} P(\alpha)=
422781389568\,\alpha+166843216800\,{
\alpha}^{2}+40063436856\,{\alpha}^{3}+6512032020\,{\alpha}^{4}+3924093
\,{\alpha}^{7}+63580545\,{\alpha}^{6}\\+753772094\,{\alpha}^{5}+{\alpha}
^{11}+105\,{\alpha}^{10}+174690\,{\alpha}^{8}+5388\,{\alpha}^{9}+
493650339840\end{array}.
\]
}
\item For $N=20$ we have (after a computation of few seconds on a standard personal computer):
{\footnotesize
\[
 \langle T_1^4T_2^3T_3^2\rangle_\alpha={Q(\alpha)(17+\alpha)(18+\alpha)(19+\alpha)\over R(\alpha)}
\]
} with {\footnotesize
\[\begin{array}{l}
Q(\alpha)=17973994269257913600\alpha+3006012942356996160\alpha^2+311313563528661024\alpha^3+22279414065220920\alpha^4\\
+1371214528697\alpha^7+\alpha^{13}+360\alpha^{12}+45753367235370\alpha^6+1164250996862956\alpha^5+\\
62879\alpha^{11}+6977370\alpha^{10}+31407665820\alpha^8+544626843\alpha^9+50279359153701888000
\end{array}
\] }

and {\footnotesize\[ R(\alpha)=
(31+\alpha)(32+\alpha)(33+\alpha)(34+\alpha)(35+\alpha)(36+\alpha)(37+\alpha)(38+\alpha)^2(39+\alpha)^3(40+\alpha)^2(41+\alpha)(42+\alpha).
\]}
\item For $N=3\dots 50$, see Fig. \ref{beta2lambda432}.
\end{itemize}
\begin{figure}[h]
\begin{center}
\resizebox{8cm}{7cm} {\includegraphics{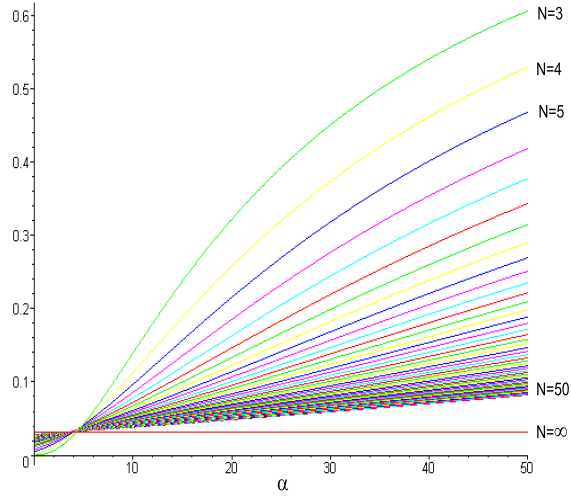}}
\caption{\label{beta2lambda432} Values of  $\langle
T_1^4T_2^3T_3^2\rangle_\alpha$ as a function of $\alpha\in[0,50]$ for $\beta=2$. From \eqref{factorizationlimits} and \eqref{caseslimit} for $p<1$,
one has $\lim_{N\to\infty} \langle
T_1^4T_2^3T_3^2\rangle_\alpha = 525/16384\simeq 0.032$ in good agreement with the plot.}
\end{center}\end{figure}
\end{Example}

{\bf Asymptotic analysis of the unitary case for $N\to\infty$.} This analysis is reported in Appendix \ref{AppD}.
\subsection{The Orthogonal ($\beta=1$) and Symplectic case ($\beta=4$)}
In complete analogy to the case $\beta=2$, the algorithm to
compute (very quickly) $\langle T_1^{\lambda_1}\dots
T_N^{\lambda_N}\rangle_\alpha$ from \eqref{quantity} consists of three steps:
\begin{enumerate}
\item Replace the permanent $(1/N!)\mathrm{perm}(T_i^{\lambda_j})_{1\leq i,j\leq N}=T_1^{\lambda_1}\cdots T_N^{\lambda_N}$ with a monomial symmetric function using \eqref{identityperm};
\item Expand the
monomial function $m_\lambda$ in the Jack basis for the parameter
$c=2$ ($\beta=4$) or $c=1/2$ ($\beta=1$) \cite{macdonald};
\item Replace each occurence of $J^{(\frac{2}{\beta})}_\mu$ by
${\lambda^{!}\over N!}{Z_{\omega\equiv J}}(\beta,N)^{-1}I^{(\frac{2}{\beta})}_\mu$
\end{enumerate}
Let us provide a couple of examples for such procedure:
\begin{Example}

Consider the average $\langle T_1^{4}T_2^{3}T_3^2 \rangle_\alpha$
for $\beta=4$. The expansion of the monomial function
$m_{[4,3,2]}$ in the Jack basis is (for an alphabet of size $N=3$)
\begin{equation}
 m_{[4,3,2]}= -{2\over 1575}J^{(\frac12)}_{[3,3,3]}+{4\over 2025}J^{(\frac
 12)}_{[4,3,2]}.
\end{equation}
After substitutions, one obtains:
 \begin{equation}
\langle T_1^{4}T_2^{3}T_3^2 \rangle_\alpha=-{2\over
1575}{I^{(\frac{1}{2})}_{[3,3,3]}\ \over 6\ Z_{\omega\equiv J}(4,3)}+{4\over 2025}{I^{(\frac{1}{2})}_{[4,3,2]}\ \over
6\ Z_{\omega\equiv J}(4,3)},
\end{equation}
which simplifies to:
\begin{equation}
\langle T_1^4T_2^3T_3^2\rangle_\alpha={\frac {\left
(59+14\,\alpha+{\alpha}^{2}\right )\left (4+\alpha
\right )^{2}\alpha\,\left (1+\alpha\right )\left (2+\alpha\right )
\left (3+\alpha\right )}{\left (7+\alpha\right )^{2}\left (8+\alpha
\right )\left (9+\alpha\right )^{2}\left (10+\alpha\right )\left (11+
\alpha\right )\left (12+\alpha\right )}}
\end{equation}

\begin{figure}[h]
\begin{center}
\resizebox{8cm}{7cm} {\includegraphics{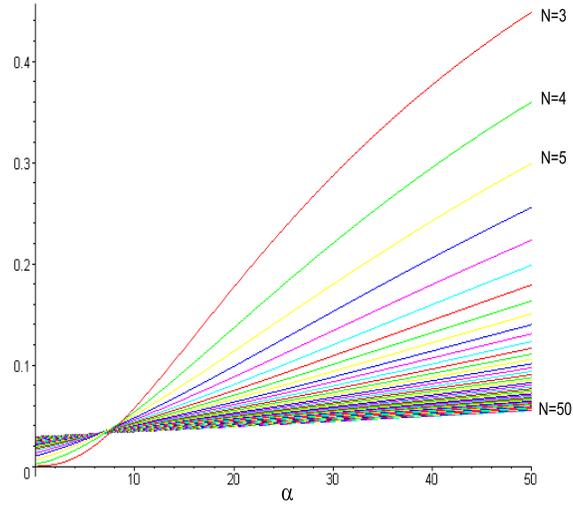}}
\caption{\label{beta4lambda432} Values of  $\langle
T_1^4T_2^3T_3^2\rangle_\alpha$ as a function of $\alpha\in [0,50]$ for $\beta=4$.}
\end{center}\end{figure}
See Fig. \ref{beta4lambda432} for other examples of evaluation of
$\langle T_1^{4}T_2^{3}T_3^2 \rangle_\alpha$. These evaluations
have taken a few seconds on a standard laptop.

\end{Example}

\begin{Example}
Consider the same average $\langle T_1^{4}T_2^{3}T_3^2
\rangle_\alpha$ for $\beta=1$. The expansion of the
monomial function $m_{[4,3,2]}$ in the Jack basis is (for an alphabet
of size $N=3$)
\[
m_{[4,3,2]}=-{\frac {1}{50400}}\, J^{(2)}_{[3,3,3]}+{\frac
{1}{18144}}\, J^{(2)}_{ [4,3,2]}
\]
After substitutions, one obtains:
 \begin{equation}
\langle T_1^{4}T_2^{3}T_3^2 \rangle_\alpha=-{\frac
{1}{50400}}{I^{(2)}_{[3,3,3]}\ \over 6\ Z_{\omega\equiv J}(1,3)}+{\frac
{1}{18144}}{I^{(2)}_{[4,3,2]}\ \over 6\ Z_{\omega\equiv J}(1,3)},
\end{equation}
which simplifies to
\begin{equation}
\langle T_1^4T_2^3T_3^2\rangle_\alpha={\frac {\left (17+8\,\alpha+{\alpha}^{2}\right
)\alpha\,\left (1+
\alpha\right )^{2}\left (1+2\,\alpha\right )\left (3+2\,\alpha\right )
}{\left (6+\alpha\right )\left (5+\alpha\right )\left (3+\alpha\right
)\left (4+\alpha\right )^{2}\left (2\,\alpha+7\right )\left (9+2\,
\alpha\right )}}
\end{equation}

\begin{figure}[h]
\begin{center}
\resizebox{8cm}{7cm} {\includegraphics{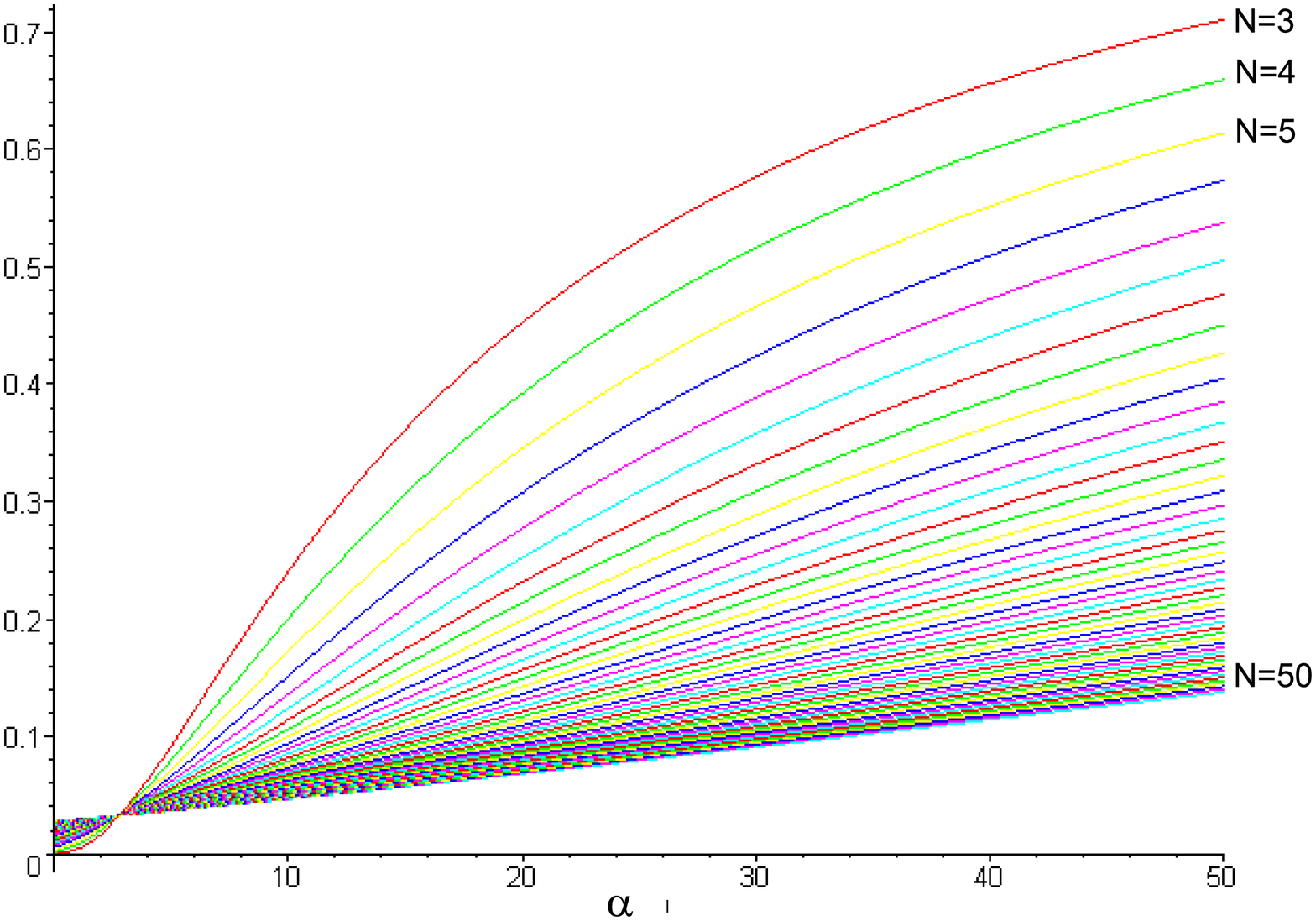}}
\caption{\label{beta1lambda432} Values of  $\langle
T_1^4T_2^3T_3^2\rangle_\alpha$ as a function of $\alpha\in [0,50]$ for $\beta=1$.}
\end{center}\end{figure}
See Fig. \ref{beta1lambda432} for other examples of evaluation of
$\langle T_1^{4}T_2^{3}T_3^2 \rangle_\alpha$.
\end{Example}
{\bf Asymptotic analysis for $N\rightarrow\infty$}: The
computations (see {\it e.g.} Fig. \ref{beta4lambda432} and
\ref{beta1lambda432}) and eq. \eqref{limitbeta4} suggest  asymptotic behaviors similar to the
case $\beta=2$. We will detail this in a forthcoming paper \cite{vivopreparation}.

\section{Conclusions}\label{conclusions}
In this paper we considered nonlinear statistics of permanents on the eigenvalues of classical invariant random matrix ensembles. Motivated by applications to the problem of quantum transport in
chaotic cavities, we first gave general formulae (based on a hyperdeterminant version of the Heine identity) for averages of permanents, valid for $\beta=2$ , any weight function $\omega(x)$ and any set of
permanent functions $\{\psi_i(x)\}$. The question of numerical efficiency is then addressed, and much quicker algorithms are found for the specific case of Jacobi weight with $\gamma=1$: the analysis is based on symmetric functions expansions, whose main merit being that the complexity does not grow at all with $N$ (the number of integration variables). This results first in a remarkable increase in efficiency, and secondly it assists performing an asymptotic analysis for large $N$. This reveals an interesting combinatorial structure lurking behind, and a detailed analysis of the \emph{factorization} conjecture we put forward is deferred to a separate publication \cite{vivopreparation}.

\begin{acknowledgements}
PV acknowledges useful discussions with G. Akemann, M.J. Phillips, D.V. Savin and R. Santachiara. JGL acknowledges A. Lascoux and C. Carr\'e for fruitful discussions.
JGL is partially supported by the ANR project PhysComb (BLAN/2008/PhysComb).

\end{acknowledgements}

\appendix
\section{Quantum transport in chaotic cavities: the random scattering matrix approach}\label{AppA}

Consider an open chaotic
cavity of sub-micron dimensions with $N_1$ and $N_2$ electronic channels in the two attached
leads. Once the system is brought out of equilibrium by an applied voltage, it is well established that the electrical current flowing
through such a cavity displays time-dependent
fluctuations, associated with the granularity of
the electron charge $e$, which persist down to zero temperature
\cite{beenakker}.

We consider here the Landauer-B\"{u}ttiker scattering approach
\cite{beenakker,landauer,buttikerPRL}. This amounts to relating the wave function
coefficients of the incoming and outgoing electrons
through the unitary scattering matrix $S$ ($2\hat{N}\times 2\hat{N}$, if $\hat{N}=N_1+N_2$):
\begin{equation}\label{ScatteringMatrix S}
  S=
  \begin{pmatrix}
    r & t^\prime \\
    t & r^\prime
  \end{pmatrix}
\end{equation}
where ($t,t^\prime$) and
$(r,r^\prime)$ stand for transmission and reflection submatrices among different channels.

The theory predicts that many interesting experimental quantities are represented by linear statistics (see \eqref{linearstat}) on
the eigenvalues of the $N\times N$ hermitian matrix $t t^\dagger$ (if $N=\min(N_1,N_2)$): for
example, the dimensionless conductance and the shot noise are
given respectively by $\mathcal{G}=\Tr(t t^\dagger)$ \cite{landauer} and
$\mathcal{P}=\Tr[t t^\dagger(1-t t^\dagger)]$ \cite{lesovik,ya}.

Random Matrix Theory, along with insightful semiclassical approaches \cite{richter,braun,berkolaiko,schanz,whitney}
is known to be very effective in describing universal fluctuation statistics in open cavities.
The simplest assumption is that the scattering matrix $S$ for the case
of chaotic dynamics is drawn from a suitable
ensemble of random unitary matrices \cite{muttalib,stone,mellopereira,altshuler}.
Assuming then ballistic point contacts \cite{beenakker}, a maximum entropy approach leads the
probability distribution of
$S$ to be uniform within the unitary group, i.e. $S$ belongs to
one of Dyson's Circular Ensembles \cite{mehtabook,Dys:new}.

The unitarity constraint induces a certain joint probability density on the transmission eigenvalues $\{T_i\}$ of the
matrix $t t^\dagger $, from which the statistics of interesting experimental quantities could be in principle derived.
This jpd is exactly of the Jacobi form with $\gamma=1$ considered throughout this paper \cite{mellopereira,beenakker,forrcond}:
\begin{equation}\label{jpd transmission First}
  P(T_1,\ldots, T_N)=\frac{1}{Z_{\omega\equiv J}(\beta,N)}\prod_{j<k}|T_j-T_k|^\beta\prod_{i=1}^N
  T_i^{\alpha-1}
\end{equation}
where the Dyson index $\beta$ characterizes different symmetry
classes ($\beta=1,2$ according to the presence or absence of
time-reversal symmetry and $\beta=4$ in case of spin-flip
symmetry), and
\begin{equation}\label{alphaN}
\alpha = \frac{\beta}{2}(|N_1-N_2|+1)
\end{equation}

The eigenvalues $T_i$ are thus correlated random variables
between $0$ and $1$ and have an intuitive interpretation in terms of the probability that an electron gets transmitted through the $i$th channel. From \eqref{alphaN},
assuming $N_1 = \ell N_2$ one has $\alpha(N)\sim (\beta/2)(\ell-1) N$ for large $N$ (see eq. \eqref{alfalinear}).

From \eqref{jpd transmission First}, in principle the statistics of all the
interesting quantities can be tackled, such as:
\begin{align}\label{List}
  \mathcal{G} &= \sum_{i=1}^N T_i &\text{(conductance)}\\
  \mathcal{P} &= \sum_{i=1}^N T_i(1-T_i)  &\text{(shot noise)}\\
  \mathcal{T}_p &= \sum_{i=1}^N T_i^p  &\text{(integer moments)}
\end{align}
along with any other linear statistics
$\mathcal{A}=\sum_{i=1}^N f(T_i)$. \\
\\
{\bf Linear and nonlinear statistics.}
The average and variance of the
above quantities are known, both for large $N$
\cite{beenakker,brouwer,novaes} and, very recently, also for a
fixed and finite number of channels $N_1,N_2$
\cite{vivovivo,savin,savin2}.
The \emph{full}
distribution of the quantities above has started recently to be the subject of thorough investigations: for the conductance, a
formula was derived for $N_1=N_2=1,2$
\cite{baranger,jalabert,garcia}. Until very recently, the distribution of the shot noise was known only for $N_1=N_2=1$ \cite{pedersen}.
Then, in \cite{sommers,savinnew}
formulas for the distribution of conductance and
shot noise, valid at arbitrary number of open channels and for any $\beta$, are derived, among other many interesting results.
In \cite{Kanz} and \cite{Kanz2}, recursion formulas for the efficient computation of conductance and shot noise cumulants were reported.
In a recent letter \cite{vivoPRL}, a large deviation approach to the problem of finding full distribution
of linear statistics in chaotic cavities (valid for a large number of open channels in the two leads) was put forward.
At odds with the linear statistics, about which virtually everything is known, the nonlinear statistics mainly considered in this paper $\langle T_1^{\lambda_1}\cdots T_N^{\lambda_N}\rangle_\alpha$
is more tricky. It appears naturally when considering moments of linear
statistics such as $\langle \mathcal{G}^n\rangle_\alpha$ or $\langle \mathcal{P}^n\rangle_\alpha$, as well as covariances of linear statistics such as $\mathrm{cov}(\mathcal{G},\mathcal{P})$, after
expanding the above mentioned averages using the multinomial theorem. Results about these objects have recently appeared \cite{savin,savinnew,novaes} and the present work is yet another step in the same direction.

\section{Hyperdeterminants}\label{AppB}

The notion of hyperdeterminant was first defined by A. Cayley in
1843 during a lecture at the Cambridge Philosophical Society,
about the possibility of extending the notion of determinant to
higher dimensional arrays. The simplest generalization is given
for a $k$th order tensor on an $n$-dimensional space
$M=(M_{i_1,\ldots, i_k})_{1\leq i_1,\ldots, i_k\leq n}$ as:
\begin{equation}\label{defhyper}
\Det\ M = \frac{1}{n!}\sum_{\sigma =(\sigma_1,\ldots,\sigma_k)\in\S_n^k}\epsilon(\sigma) M^\sigma
\end{equation}
where $\epsilon(\sigma)$ is the product of signatures of the $k$ permutations, $M^\sigma=M_{\sigma_1(1),\ldots,\sigma_k(1)}\cdots M_{\sigma_1(n),\ldots, \sigma_k(n)}$
and $\S_n$ is the symmetric group. It is straightforward to see that $\Det\ M =0$ if $k$ is odd.

A further refinement is due to L. Gegenbauer (see {\it
e.g. }\cite{lecat}) in 1890 who generalized \eqref{defhyper} to the
case where some of the indices are non-alternated. More precisely,
if $I$ denotes a subset of $\{1,\ldots,k\}$ one has:
\begin{equation}\label{defhyper2}
\Det_I\ M = \frac{1}{n!}\sum_{\sigma =(\sigma_1,\ldots,\sigma_k)\in\S_n^k}\epsilon\left(\prod_{i\in I}\sigma_i\right) M^\sigma
\end{equation}

In particular, in the main text we defined:
\begin{equation}\label{defdetplus}
\Det_+(M_{i_1,i_2,i_3})=\Det_{\{2,3\}}(M_{i_1,i_2,i_3})
\end{equation}

No matter how many indices are non-alternated, every hyperdeterminant admits an expansion in sums of lower-order hyperdeterminants. More precisely,
each hyperdeterminant of dimension $k$ (where the dimension is just the number of indices) and order $n$
 is equal to a linear combination of $(n!)^{k-\ell}$ hyperdeterminants of dimension $\ell$ and order $p$. The hyperdeterminants in the sum are obtained by fixing some of the indices.
 In particular, it is always possible to expand a Gegenbauer hyperdeterminant (see {\it e.g.} \cite{lecat}) as a sum of $(n!)^{k-2}$ conventional determinants:
 \begin{equation}\label{GegenExp}
 \Det_I\ M = \sum_{\sigma_3,\ldots,\sigma_k\in\S_n^{k-2}}
 \epsilon\left(\prod_{i\in I}\sigma_i\right) \det(M^{\sigma_3,\ldots,\sigma_k})
 \end{equation}
 where $M^{\sigma_3,\ldots,\sigma_k}$ denotes the $n\times n$ matrix such that $(M^{\sigma_3,\ldots,\sigma_k})_{i,j}$ = $M_{i,j,\sigma_3(i),\ldots,\sigma_k(i)}$.

 Combining \eqref{GegenExp} with \eqref{defdetplus}, one easily obtains the expansion in \eqref{eq2}.
Note that a more general version of \eqref{GegenExp} can be found
in \cite{luque}.

\section{Partitions and symmetric functions}\label{AppC}
\def\X{{\mathbb X}}
A {\it partition} is a finite sequence $\lambda=(\lambda_1, \lambda_2, \ldots , \lambda_N)$ of non-negative integers (called parts) such that 
$\lambda_1 \ge \lambda_2 \ge \ldots \ge \lambda_N \ge 0$. We define the {\it weight} of a partition $|\lambda|$ as the sum of its parts, and its {\it length}, $\ell(\lambda)$ as the number of its non-zero parts. Two partitions differing only by the number of their zero parts coincide. One can think of unidentical partitions of weight $N$ as different ways to write the integer $N$ as
sums of positive integers. For example, one has only one partition $\lambda=(1)$ in the case of $N=1$, but two partitions $\lambda = (2,0),(1,1)$ for $N=2$ and three $\lambda = (3,0,0), (2,1,0), (1,1,1)$ for $N=3$.

The {\it symmetric
functions} are polynomials in several variables
$\X=\{x_1,\dots,x_n,\dots\}$ which are invariant by permutation of
the variables.
 The set of all these polynomials for a given alphabet is an
 algebra $\Lambda$. 
 In the case where there is no relation between the
 variables (this implies in particular that the alphabet is
 infinite), the elements of the bases of the space $\Lambda$ are
 indexed by partitions. This is the case, for instance, for the
 monomial functions which are defined by
 \begin{equation}
 m_\lambda(\X)=\frac1{\lambda^!}\sum_{i_1,\dots,i_k}
 x_{i_1}^{\lambda_1}\dots x_{i_k}^{\lambda_k}
 \end{equation}
 where the already defined symbol $\lambda^!=\prod_i j_i!$ if $\lambda=[\lambda_1\geq
 \lambda_2\geq\dots\geq\lambda_k]=[\dots i^{j_i}\dots
 2^{j_2}1^{j_1}]$ and  $\lambda_k>0$.
Now, if we orthogonalize this basis w.r.t. the standard scalar
product over the symmetric functions, we obtain the basis of Schur
functions $s_\lambda$. The Gram-Schmidt 
algorithm allows to write monomial functions as a linear
combination of Schur functions. For a given partition $\lambda$, the Schur polynomial is defined as:
\begin{equation}\label{schur}
s_{\lambda}(x_1, \ldots , x_n)=\frac{\det \left (
x_i^{\lambda_j+n-j}\right )_{1\leq i,j\leq n}}{\det\left (
x_i^{n-j}\right )_{1\leq i,j\leq n}}
\end{equation}

The denominator in \eqref{schur} is the Vandermonde determinant $\prod_{i<j}(x_i-x_j)$. For partitions composed by just one part, $\lambda = (r)$, Schur functions are just the complete symmetric functions, $s_{(r)}(x)=h_r$ \cite{macdonald}, while for partitions of the form $\lambda =(1,\ldots, 1)\equiv (1^r)$, the Schur functions $s_{(1^r)}$ are the elementary symmetric functions $e_r(x)$. Schur functions 
corresponding to partitions of $N$ form a basis in the space of homogeneous symmetric polynomials of degree $N$, so that any homogeneous symmetric polynomial 
can be written as a linear combination of Schur functions.

There is an efficient way to compute such expansions.
Let us first see an example:
\begin{Example}
Suppose that we want to compute the Schur expansion of
$m_{[3,1]}$. We consider the alphabet $\X=\{x_1,x_2,x_3,x_4\}$.
Evaluated on $\X$, the monomial function gives
\[
\begin{array}{rcl}
m_{[3,1]}(\X)&=&x^{3100}+x^{3010}+x^{3001}+x^{0310}+x^{0301}+x^{0031}+\\
&&x^{1300}+x^{1030}+x^{1003}+x^{0130}+x^{0103}+x^{0013},
\end{array}
\]
where $x^{i_1i_2i_3i_4}=x_1^{i_1}x_2^{i_2}x_3^{i_3}x_4^{i_4}$.
Hence,
\[
\begin{array}{rcl}
m_{[3,1]}(\X)&=&s_{3100}+s_{3010}+s_{3001}+s_{0310}+s_{0301}+s_{0031}+\\
&&s_{1300}+s_{1030}+s_{1003}+s_{0130}+s_{0103}+s_{0013}\\
&=& s_{3100} + 0 + 0 - s_{211}+0 + s_{1111}+\\
&&-s_{2200} + 0 + s_{1111}+0+0+0\\
&=& s_{31}-s_{211}-s_{22}+2s_{1111}.
\end{array}
\]
\end{Example}

The standard algorithm thus goes as follows (see exercise 11 pag. 
110 in \cite{macdonald}):
\begin{enumerate}
\item Expand the monomial symmetric function $m_\lambda$ in the variables $x^I$
\begin{equation}
m_\lambda = \sum x^I
\end{equation}
where $I$ stands for all distinct permutations of $\lambda$ considered as a 
vector of size $N$, completed by $0$s if necessary (ex: $[3,1]\sim[3,1,0,0]$ 
for $N=4$), and $x^I=x_1^{I_1}\dots x_N^{I_N}$.
\item Replace each $x_I$ by a generalized Schur function $s_I$, defined as:
\begin{equation}
s_I=\frac{\det(x_i^{I_j+n-j})_{1\leq i,j\leq n}}{\prod_{i<j}(x_i-x_j)}=\det(s_{[I_i-i+j]})_{1\leq i,j\leq \ell(I)}
\end{equation}
where $s_0=1$ and $s_{-i}=0$ for each $i>0$.
Note that such generalized Schur function is equal to a traditional Schur function times a coefficient $0$ or $\pm 1$.
\item Replace each $s_I$ by $0,\pm 1$ times the corresponding Schur function, according to the rule
for $i<j$:
\begin{equation}
 s_{\dots,i,j,\dots}=\left\{\begin{array}{ll}-s_{\dots,j-1,i+1,\dots}&\mbox{ if } i<j-1\\
 0&\mbox{ if } i=j-1 \end{array}\right.
\end{equation}
\end{enumerate}

Another important example of basis is given by the Jack
polynomials which are a one-parameter deformation of the Schur
functions. We follow the notation of \cite{macdonald}. One starts from the deformation of the usual scalar
product defined on power sums by
\begin{equation}
 \langle p_\lambda,p_\mu\rangle_\xi =
 \xi^{\ell(\lambda)}z_\lambda\delta_{\lambda,\mu}.
\end{equation}
where $p_{[\lambda_1,\dots,\lambda_k]}=p_{\lambda_1}\dots
p_{\lambda_k}$ and $p_n=\sum_{x\in\X} x^n$. The coefficient $z_\lambda$ is given by:
\begin{equation}
z_\lambda =\prod_{i=1}^{\ell(\lambda)} a_i!\ i^{a_i},
\end{equation}
$a_i$ being the number of occurrences of $i$ in $\lambda$.

 The Jack basis
$P_\lambda^{(\xi)}$ is obtained orthogonalizing the monomial
basis with respect to the dominance order $\prec$. This means:
\begin{enumerate}
\item $\langle P_\lambda^{(\xi)},P_\mu^{(\xi)}\rangle_\xi=0$ if $\lambda\neq\mu$
\item $P_\lambda^{(\xi)}=\sum_{\mu\prec\lambda}v_{\lambda\mu}(\xi)m_\mu$
\end{enumerate}
where $\mu\prec\lambda$ means $\sum_{i=1}^\kappa \mu_\kappa\leq\sum_{i=1}^\kappa\lambda_i$ for all $\kappa$.

\begin{Example}
  One has
  \[
 \begin{array}{rcl}
    m_{[1,1,1]}&=&P_{[1,1,1]}^{(\xi)},\\
    m_{[2,1]}&=&P_{[2,1]}^{(\xi)}+{\langle m_{[2,1]},P_{[1,1,1]}^{(\xi)}\rangle_\xi\over \langle P_{[1,1,1]}^{(\xi)},P_{[1,1,1]}^{(\xi)}\rangle_\xi
    }P_{[1,1,1]}^{(\xi)}\\
    &=&P_{[2,1]}^{(\xi)}-\frac {6}{\xi+2}P_{[1,1,1]}^{(\xi)}\\
   m_{[3]}&=&P_{[3]}^{(\xi)}+{\langle m_{[3]},P_{[2,1]}^{(\xi)}\rangle_\xi\over \langle P_{[2,1]}^{(\xi)},P_{[2,1]}^{(\xi)}\rangle_\xi
    }P_{[2,1]}^{(\xi)}+{\langle m_{[3]},P_{[1,1,1]}^{(\xi)}\rangle_\xi\over \langle P_{[1,1,1]}^{(\xi)},P_{[1,1,1]}^{(\xi)}\rangle_\xi
    }P_{[1,1,1]}^{(\xi)}\\
    &=&P_{[3]}^{(\xi)}-\frac3{2\,\xi+1}\,
{ P_{[2,1]}^{(\xi)}} +\frac {6}{\left
(\xi+2\right )\left (\xi+1\right )}P_{[1,1,1]}^{(\xi)}
 \end{array}
  \]
\end{Example}
Unlike the Schur functions, the Jack polynomials
$P_\lambda^{(\xi)}$ are not orthonormal. Many normalizations are
encountered in literature. The normalization $J_\lambda^{(\xi)}$
which is used in our paper is the integral form of
$P_\lambda^{(\xi)}$:
\[
 J_\lambda^{(\xi)}=\prod_{s\in\lambda}(\xi a_\lambda(s)+\ell_\lambda(s)+1)
 P_\lambda^{(\xi)}
\]
where the product is over the nodes $s=(i,j)$ of the partitions
$\lambda$ (regarded as a tableaux), $a_\lambda(s)=\lambda_i-j$ and
$\ell_\lambda(s)={\lambda'}_j-i$ if $\lambda'$ denotes the
conjugate partition of $\lambda$.

\section{Asymptotic Analysis for $N\to\infty$ and $\beta=2$}\label{AppD}
The main formula \eqref{mainjacobibeta2} lends itself to a quite
interesting asymptotic analysis for $N\to\infty$. Since the sum
does not depend on $N$, one may be tempted to analyze the large
$N$ asymptotics of individual summands. Quite interestingly, this
is not sufficient: the individual summands actually diverge when
$N\to\infty$, whereas the full $\mu$-sum converges as it
should. More precisely, each individual summand factorizes into
the product of $\Diamond$) a convergent term depending of $\alpha$
and $\Box)$ a divergent term with a polynomial asymptotic behavior
but with no $\alpha$-dependence. Indeed, if one replaces the
normalization constant by its explicit value, one can cast
\eqref{mainjacobibeta2} in the form:
\begin{equation}\label{averagereplaced}
\langle T_1^{\lambda_1}\cdots T_M^{\lambda_M}\rangle_\alpha=\lambda^!
\sum_\mu \tK_\lambda^\mu \underbrace{\prod_{i<j}{\mu_i-\mu_j+j-i\over
j-i+1}}_{(\Box)}\ \underbrace{\prod_{i=1}^{M}\prod_{j=1}^{\mu_i}{\mu_i+N-i-j+\alpha\over
2N+\mu_i-i-j+\alpha}}_{(\Diamond)}
\end{equation}
where $M=\ell(\lambda)\leq N$ does not increase with $N$.

No matter what the dependence of $\alpha$ on $N$ is, the fact that $(\Diamond)$ converges for $N\to\infty$ is evident being the ratio of
polynomials in $N$ of the same order.

We are interested in computing the following limit:
\begin{equation}
\boxed{\mathcal{L}_\lambda=\lim_{N\to\infty} \langle T_1^{\lambda_1}\cdots T_M^{\lambda_M}\rangle_{\alpha(N)}}
\end{equation}
where the dependence of $\alpha$ on $N$ is arbitrary.
In the context of the present study (see Appendix \ref{AppA}), the parameter $\alpha$ is a linear function in
$N$
\begin{equation}\label{alfalinear}
 \alpha(N)={\beta\over 2}(\ell-1)N+1,\, \qquad (\mbox{here }\beta=2).
\end{equation}
but one can extend the results to the cases where $\alpha$ is a
polynomial in $N$,
$$\alpha(N)=(\ell-1)N^p+\sum_{m<p}b_mN^m,$$
with $p\in\Q$.  Note that only the highest degree part of
$\alpha(N)$ gives contribution in the limit $N\to\infty$.
Hence, we will only consider the case where
$$\alpha(N)\sim (\ell-1)N^p$$ with $p\in\Q$.
 The computation of the sought asymptotics is now straightforward using symbolic softwares, as in the following example.

\begin{Example}
Suppose one has to compute $\mathcal{L}_{[2]}=\lim_{N\rightarrow \infty}
\langle T_1^{2}\rangle_{\alpha}$. First, using
(\ref{mainjacobibeta2}), one obtains after
simplification of each summand: \begin{equation}
\langle
T_1^{2}\rangle_{\alpha}=\frac12\,{\frac {\left (1+N\right )\left
(N+\alpha\right )\left (-1+N+ \alpha\right )}{\left
(2\,N+\alpha\right )\left (2\,N-1+\alpha\right )
}}-\frac12\,{\frac {\left (N-1\right )\left (-1+N+\alpha\right
)\left (-2+ N+\alpha\right )}{\left (2\,N-1+\alpha\right )\left
(2\,N-2+\alpha \right )}}.
\end{equation}
Note that each individual summand
does not converge for $N\to\infty$ as remarked above. After simplifying the full expressions one obtains:
\begin{equation}
\langle
T_1^{2}\rangle_{\alpha}= {\frac {\left (-1+N+\alpha\right )\left
(-3\,N+3\,N\alpha-2\,\alpha+{ \alpha}^{2}+3\,{N}^{2}\right
)}{\left (2\,N+\alpha\right )\left (2\,N- 1+\alpha\right )\left
(2\,N-2+\alpha\right )}}.
\end{equation}
This expression has the same degree in $N$ in the numerator and denominator, no matter what the dependence of $\alpha$ on $N$ is.
So the limit exists and is given by the ratio of the highest powers of $N$.
\end{Example}
 \begin{Example} See in Fig. \ref{beta2lambda432alphal-1}, an
example showing the convergence of $\langle
T_1^4T_2^3T_3^2\rangle_{3N+1}$.
\begin{figure}[h]
\begin{center}
\resizebox{8cm}{7cm}
{\includegraphics{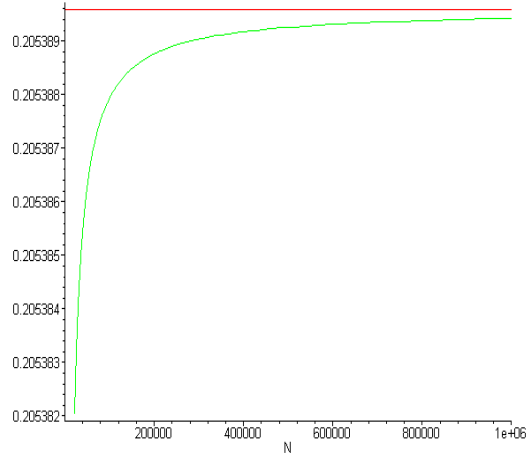}}
\caption{\label{beta2lambda432alphal-1} Values of  $\langle
T_1^4T_2^3T_3^2\rangle_{3N+1}$ as a function of $N\in [0,10^{6}]$
for $\beta=2$. From \eqref{factorizationlimits} and \eqref{caseslimit} for $p=1$ (or equivalently \eqref{novaesexpression}), one has
$\lim_{N\to\infty} \langle
T_1^4T_2^3T_3^2\rangle_{3N+1}=1253598528/6103515625\simeq 0.20539$, in full agreement with the plot.}
\end{center}\end{figure}
\end{Example}
In all cases, numerical evidences suggest the following conjecture about $\mathcal{L}_\lambda$, which will be analyzed in further detail in a forthcoming publication \cite{vivopreparation}.

\begin{conj}[Factorization of limits] \label{fact}Let $\lambda=(\lambda_1,\dots,\lambda_M)$ be a
partition. One has
\begin{equation}\label{factorizationlimits}
\mathcal{L}_\lambda=\prod_{j=1}^M\mathcal{L}_{[\lambda_j]}
\end{equation}
\end{conj}
This means that it is always sufficient to analyze the limit in the case of partitions with one single part.
\begin{Example}
Consider again the case where $\lambda=[4,3,2]$, but for
$\alpha\sim (\ell-1)N$. After a brief computation one obtains:
 \begin{equation}
\lim_{N\rightarrow \infty} \langle
T_1^{4}T_2^3T_3^2\rangle_{(\ell-1)N}=
 {\frac {\left
(1+3\,\ell+9\,{\ell}^{3}+9\,{\ell}^{2}+3\,{\ell}^{5}+9\,{\ell}^{4}+{\ell}^
{6}\right )\left (1+\ell+{\ell}^{2}\right )\left
({\ell}^{4}+2\,{\ell}^{3}+4\,{\ell}^ {2}+2\,\ell+1\right
){\ell}^{3}}{\left (1+\ell\right )^{15}}}.
\end{equation}
But one has also
\[
\begin{array}{l}\displaystyle
\lim_{N\rightarrow \infty} \langle T_1^{4}\rangle_{(\ell-1)N}=
{\frac {\ell\left
(1+3\,\ell+9\,{\ell}^{3}+9\,{\ell}^{2}+3\,{\ell}^{5}+9\,{\ell}^{4}+{\ell}
^{6}\right )}{\left (1+\ell\right )^{7}}}\\
\displaystyle\lim_{N\rightarrow \infty} \langle
T_1^{3}\rangle_{(\ell-1)N}={\frac {\ell\left
({\ell}^{4}+2\,{\ell}^{3}+4\,{\ell}^{2}+2\,\ell+1\right )}{\left
(1
+\ell\right )^{5}}}\\
\displaystyle\lim_{N\rightarrow \infty} \langle
T_1^{2}\rangle_{(\ell-1)N}={\frac {\ell\left
(1+\ell+{\ell}^{2}\right )}{\left (1+\ell\right )^{3}}}.
\end{array}
\]
Hence,
\begin{equation}
\lim_{N\rightarrow \infty} \langle
T_1^{4}T_2^3T_3^2\rangle_{(\ell-1)N}=\lim_{N\rightarrow \infty}
\langle T_1^{4}\rangle_{(\ell-1)N}\lim_{N\rightarrow \infty}
\langle T_1^{3}\rangle_{(\ell-1)N}\lim_{N\rightarrow \infty}
\langle T_1^{2}\rangle_{(\ell-1)N}.
\end{equation}
\end{Example}

Assuming Conjecture (\ref{fact}), it remains to consider the
limit
\begin{equation}\label{limitNovaes}
 \mathcal{L}_{[k]}=\lim_{N\to\infty}\hat{I}_{[k]}(\alpha,N)
 \end{equation}
where
\begin{equation}
\hat{I}_{[k]}(\alpha,N)=\langle T_1^{k}\rangle_{\alpha}
\end{equation}
  i.e. the case
when the partition $\lambda$ is composed by just one part $\lambda=[k]$.

For  $\alpha(N)\sim \frac{\beta}{2}(\ell-1)N+1$, the limit \eqref{limitNovaes} has been computed by Novaes \cite{novaes} as:
\begin{equation}\label{novaesexpression}
\mathcal{L}_{[k]}=\lim_{N\rightarrow \infty} \hat{I}_{[k]}\left(\frac{\beta}{2}(\ell-1)N,N\right)=
(\ell+1)\sum_{p=1}^k\frac{(-1)^{p-1}}{p}\binom{k-1}{p-1}\binom{2(p-1)}{p-1}\left(\frac{\ell}{(\ell+1)^2}\right)^p
\end{equation}
Guided by the numerics, we have found an equivalent expression (see \eqref{caseslimit}), whose direct combinatorial proof will be announced in a separate publication \cite{vivopreparation}.
In the case $\lambda=[k]$, the monomial function $m_{[k]}=\sum_i T_i^k$ is
the power sum $p_k$ and the coefficient $\tK_{[k],\mu}$ are well
known (see {\it e.g.} \cite{macdonald}):
\begin{equation}
 m_{[k]}=\sum_{i=0}^{k-1}(-1)^is_{[(k-i),1^i]}.
\end{equation}
Plugging these coefficients in (\ref{mainjacobibeta2}), one
recognizes, after simplification, a hypergeometric
function: \begin{equation}
\hat{I}_{[k]}(\alpha,N)={\frac {\Gamma (2\,N+\alpha-1)\Gamma
(-1+N+ \alpha+k)(N+k-1)!}{k!N!\Gamma (-1+N+\alpha) \Gamma
(2\,N+\alpha-1+k)}}\
_4F_3\left({2-N-\alpha,-2\,N-\alpha+2-k,-k+1,1-N \atop -k+1-N
,-2\,N+2-\alpha,-N-\alpha+2-k};1\right)
\end{equation}
where
\begin{equation}
 \ _pF_q\left({a_1,\dots,a_p\atop b_1,\dots,b_q};x\right)=\sum_{i\geq
 0}{(a_1)_i\dots(a_p)_i\over (b_1)_i\dots(b_q)_i}{x^i\over i!}
\end{equation}
if $(x)_i=x(x+1)\dots(x+i-1)$ denotes the rising factorial.

Suppose now that $\alpha\sim(\ell-1)N^p$. In this case, numerical evidences
suggest the following alternative representation for \eqref{novaesexpression}:

 \begin{equation}\label{caseslimit}
\mathcal{L}_{[k]}=\lim_{N\rightarrow \infty} \hat{I}_{[k]}((\ell-1)N^p,N)=
\begin{cases}
{\ell\over
(\ell+1)^{2k-1}}\sum_{i=0}^{2(k-1)}\left(k-1\atop \left\lfloor i\over
2\right\rfloor\right)\left(k-1\atop \left\lceil i\over
2\right\rceil\right)\ell^i
 &\mbox{for } p=1\\
{\left(2k-1\atop
k-1\right)\over 2^{2k-1}}&\mbox{for } p<1\\
1 &\mbox{for } p>1
\end{cases}
\end{equation}
where $\lceil\omega \rceil$ (resp. $\lfloor\omega \rfloor$)
denotes the smallest (resp. largest) integer larger (resp. smaller)
or equal to $\omega$.

 Note that:
 \begin{itemize}
 \item  The coefficient $\left(k\atop \left\lfloor i\over
2\right\rfloor\right)\left(k\atop \left\lceil i\over
2\right\rceil\right)$ has a very interesting combinatorial
interpretation, since it is also the number of symmetrical Dyck
paths with even semi-length $2k$ and exactly $i$ peaks \cite{barry}. We will
explore this
property in a forthcoming paper \cite{vivopreparation}, where a formal proof of the equivalence between 
\eqref{caseslimit} and \eqref{novaesexpression} based on the creative telescoping method will be provided.

\item Eq. \eqref{caseslimit} for $p=1$ can be equally well restated in
terms of hypergeometric functions as:

 \begin{equation}
\lim_{N\rightarrow \infty}\hat{I}_{[k]}((\ell-1)N,N)={\ell\over
(\ell+1)^{2k-1}} \left(\ _2F_1(-k,-k;1;\ell^2)+\ell k\
_2F_1(1-k,-k;2;\ell^2)\right).
\end{equation}

\item The second and third case in \eqref{caseslimit} are obtained from the $p=1$ case upon setting $\ell=1$ and $\ell\to\infty$ respectively.

\item From the factorization conjecture and the independence of Novaes' limit \eqref{novaesexpression} on the exponent $\beta$ of the Vandermonde, one 
obtains the following result. Suppose $\beta > 0$ and
set
\begin{equation}
\hat{I}_{\lambda}(\alpha,N;\beta):=\frac{1}{Z_{\omega\equiv J}(\beta,N)}\int_{[0,1]^N}d T_1\cdots
dT_N \ T_1^{\lambda_1}\cdots
T_N^{\lambda_N}\prod_{j<k}|T_j-T_k|^\beta\prod_{i=1}^N T_i
^{\alpha-1}.
\end{equation} 
One has:
\begin{equation}\label{limitbeta4}
\lim_{N\rightarrow \infty}\hat{I}_{\lambda}\left(\frac\beta2(\ell-1)N,N;\beta\right)={\cal L}_\lambda.
\end{equation}
\end{itemize}

\end{document}